\documentclass[12pt]{article}
\usepackage{amsxtra,amssymb,amsthm,amsmath,latexsym}

\theoremstyle{plain}

\newtheorem{theorem}{Theorem}[section]

\newtheorem{remark}{Remark}[section]
\numberwithin{equation}{section}

\newcommand{\refT}[1]{Theorem~\ref{T:#1}}
\newcommand{\refS}[1]{Section~\ref{S:#1}}

\def\R{{\mathbb R}}

\def\C{{\mathbb C}}

\def\oH1{{\overset{\circ}{H}\kern-.04in{}^1}}

\def\ve{{\varepsilon}}

\def\const{{\,const\,}}
\def\loc{{\,loc\,}}
\def\Fred{{\,Fred\,}}
\def\bee{\begin{equation*}}
\def\eee{\end{equation*}}
\def\be{\begin{equation}}
\def\ee{\end{equation}}

\begin{document}
\title{                  
A nonlinear singular perturbation problem}

\author{
A.G. Ramm\\
 Mathematics Department, Kansas State University, \\
 Manhattan, KS 66506-2602, USA\\
ramm@math.ksu.edu\\}

\date{}

\maketitle\thispagestyle{empty}

\begin{abstract}
\footnote{Math subject classification: 47H15, 47H17, 45G10, 
35B25;\, PACS 02.30Tb, 02.30.Rz }
\footnote{key words: nonlinear operator equations, 
bifurcation, singular perturbation, integral equations  }

Let
\bee
   F(u_\ve)+\ve(u_\ve-w)=0 \eqno{(1)}\eee where $F$ is a
nonlinear operator in a Hilbert space $H$, $w\in H$ is an
element, and $\ve>0$ is a parameter. Assume that $F(y)=0$,
and $F'(y)$ is not a boundedly invertible operator.
Sufficient conditions are given for the existence of the
solution to \eqref{e1.1} and for the convergence
$\lim_{\ve\to 0}\|u_\ve-y\|=0$. An example of applications
is considered. In this example $F$ is a nonlinear integral
operator.

\end{abstract}

\section{Introduction}

In many physical problems the behavior of a solution to an equation 
depending on a small parameter is of interest. There is a large literature 
on this topic (\cite{CH},\cite{L}, \cite{VT}). The novel point 
in
this paper is the treatment of such a problem for a nonlinear operator 
equation in a Hilbert space without the usual assumption that the 
Fr\'echet
derivative of the nonlinear operator at the solution of the limiting
equation is a Fredholm-type linear operator.

In a real Hilbert space consider a nonlinear operator $F\in C^3_\loc$,
i.e., 
$\sup_{u\in B(u_0,R)} \|F^{(j)} (u)\| \leq M_j$, $j=1,2,3$,
where $M_j=M_j(R)$ are constants, $u_0\in H$ is some 
element, $R>0$ is a number,
$F^{(j)} (u)$ are Fr\`echet derivatives,
and $B(u_0,R)=\{u:\|u-u_0\|\leq R\}$.

Assume that $F(y)=0$ and $F(u_\ve)+\ve(u_\ve-w)=0$
$\forall\ve>0$, $\ve\in(0,\ve_0)$, $\ve_0>0$, $w\in H$ is an
element. We are interested in the conditions under which
$\lim_{\ve\to 0} \|u_\ve-y\|=0$. This question has been
studied much in the literature (\cite{CH},\cite{VT}) under
the following assumptions:

i) $\exists F'(y):=A$

ii) $A$ is an isomorphism of $H$ onto $H$,

or, in place of ii) one may make a weaker assumption:

iii) $A\in \Fred(H)$, i.e., $A$ is a Fredholm-type operator,

that is, the range $R(A)$ is closed,
the null-space $N(A)$ is finite-dimensional, $\dim N(A)=n<\infty$,
and $\dim N(A^\ast)=n^\ast<\infty$.

One may define
$z_\ve:=u_\ve-y$, $z_0=0$, $F(y+z_\ve):=\phi(z_\ve)$,
and consider the following equation:
\be\label{e1.1}
  \phi (z_\ve)+\ve z_\ve+\ve(y-w)=0.\ee
The problem is to prove (under suitable assumptions) that
\be\label{e1.2}
  \lim_{\ve\to 0} \|z_\ve\|=0. \ee 
One has
$A:=F'(y)=\phi'(0)$. If the above assumptions i) and ii) 
hold,
then it is known (see \cite{VT}) that equation \eqref{e1.1}
has a unique solution $z_\ve$ for all
$\ve\in(0,\ve_0)$,where $\ve_0>0$ is sufficiently small, and
\eqref{e1.2} holds. Indeed, using the Taylor formula, one gets:
\be\label{e1.3}
  \phi (z_\ve)=Az_\ve+K(z_\ve),
  \quad \|K(z_\ve)\|\leq \frac{M_2\|z_\ve\|^2}{2}.\ee
One writes \eqref{e1.1} as
\be\label{e1.4}
  z_\ve=-A^{-1}_\ve K(z_\ve)-\ve A^{-1}_\ve(y-w),\quad
A_\ve:=A+\ve I,\ee 
and applies the contraction mapping theorem
 to \eqref{e1.4}. This yields the existence of a
 solution $z_\ve$ to \eqref{e1.1} for all
$\ve\in(0,\ve_0)$ and the convergence result \eqref{e1.2}
in a stronger form $||z_\ve||=O(\ve)$. The application
of the  contraction mapping theorem is possible because of 
the estimate $\sup_{\ve\in (0,\ve_0}||A_\ve^{-1}||\leq c$,
which is easy to prove if $||A^{-1}||\leq c$, that is, if 
assumptions i) and ii) hold. 

If $A\in\Fred(H)$ then there is also a bifurcation theory
for equation \eqref{e1.1}, but it is more complicated than
in the case when $A$ is an isomorphism of $H$ onto $H$ (see
\cite{VT} and \cite{CH}).

\textit{The main novel point of our paper is a study of
equation \eqref{e1.1} in the case when $A$ is not a
Fredholm-type operator.}

For example,  $A$ can be a compact operator.

Our basic result is a proof of relation \eqref{e1.2} under
the assumption \be\label{e1.5}
  \|(A+\ve)^{-1}\|\leq c\ve^{-1}, \quad \ve>0, c=\const, \ee
where $c$ does not depend on $\ve$, and $w$ is suitably chosen.

Condition \eqref{e1.5} holds if there exists a set
$\{\zeta\in \C:|\zeta|<r$, $\pi-\alpha<\arg 
\zeta<\pi+\alpha\}$ which
consists of regular points of the operator $A:=F'(y)$. Here
$r>0$ and $\alpha>0$ are arbitrary small positive numbers.
In particular, if $A\geq 0$ is a selfadjoint operator, then
\eqref{e1.5} holds with $c=1$.

As an application of Theorem 2.1, Section 2,  we study the 
following integral equation:
\be\label{e1.6}
  \int_D g(x,s)u^3_\ve(s)ds+\ve(u_\ve-w)=f,\quad
g(x,s):=(4\pi|x-s|)^{-1}, \quad f\in C^\infty_0(D).
\ee 
Under suitable assumptions on
$f$ we prove \eqref{e1.2} with $y$ being a solution to
equation \be\label{e1.7}
  \int_D\frac{y^3(s)ds}{4\pi|x-s|}=f.\ee

In \refS{2} Theorem 2.1 is formulated and proved.

In \refS{3} equation \eqref{e1.6} is studied.

\section{Formulation and proof of the result}\label{S:2}
\begin{theorem}\label{T:2.1}

Assume that $F\in C^3_\loc$, $F(y)=0$, $F$ is compact,  \eqref{e1.5} 
holds with
$A:=F'(y)$, and $w$ is such that $y-w=Av$, where
$\|v\|<\frac{1}{2M_2c(1+c)}$, and $c>0$ is the constant 
in \eqref{e1.5}. Then \eqref{e1.1} has a  solution for 
all $\ve\in (0,\ve_0)$, where $\ve_0$ is sufficiently small, 
 \eqref{e1.2} holds, and $||z_\ve||=O(\ve)||$.
The solution to \eqref{e1.1} is unique in a sufficiently small
ball $\{u: ||u-y||\leq R\}$, $R=O(\ve)$.
\end{theorem}
              
\begin{remark}\label{R:2.2} If $\overline{R(A)}=H$, then one
can always find a $w$ such that $u-w=Av$ and $\|v\|<b$,
where $b>0$ is an arbitrary small number. This conclusion holds under much 
weaker assumption. Namely,
let $A_b$ denote the restriction of $A$ to the ball $B(0,b):
=\{u: ||u||\leq b\}$, where $b=const>0$ is the radius of the ball. Let
$R_b:=\{v: v=Au, u\in B(0,b)\}$, and let $\overline{R_b}$ be its closure.
If
$\overline{R_b}\cap \{B(0,r\}\setminus \{0\}\}\neq \emptyset$, where $r>0$ 
is some number, then there exists a $w$ such that
$y-w=Av$, $\|v\|\leq b$. \end{remark}

\begin{proof} Rewrite equation \eqref{e1.1} as \eqref{e1.4}.
Denote the right-hand side of \eqref{e1.4} by $T(z_\ve)$.
Let $B(R):=\{z:\|z\|\leq R\}$. Choose a suitable 
dependence $R=R(\ve)\to 0$ as $\ve\to 0$ (see \eqref{e2.2}).
Then 
$TB(R)\subset B(R)$ if
$\ve$ is sufficiently small.

Indeed,
$$A^{-1}_\ve(y-w)=A^{-1}_\ve Av=v-\ve A^{-1}_\ve v,$$
 so
$$\|A^{-1}_\ve(y-w)\| \leq \|v\|+c\|v\|,$$
 and 
\be\label{e2.1}
 \|T(z)\|\leq \|-A^{-1}_\ve K(z)-\ve A^{-1}_\ve (y-w)\|
   \leq \frac{c}{\ve} \frac{M_2 R^2}{2}+\ve\|v\|(1+c)\leq R, \ee
provided that 
\be\label{e2.2}
\frac{\ve}{cM_2}(1-\rho)\leq R\leq \frac{\ve}{cM_2}(1+\rho),
\ee
where $\rho =\sqrt{1-2M_2\|v\|c(1+c)}$.

Since $F$ is compact, so are $K$ and $T:=-A_\ve^{-1}K-\ve 
A_\ve^{-1}(y-w)$. As we have proved above, the operator $T$ maps the ball 
$B(0,R)$ into itself if $\ve$ is sufficiently small.
Therefore, by the Schauder's fixed-point theorem, the map $T$
has a fixed point in $B(0,R)$, i.e., equation \eqref{e1.4} has a solution 
in 
$B(0,R)$. Since $R=O(\ve)$, it follows that 
$||z_\ve||=O(\ve)$, so \eqref{e1.2} holds.

To prove uniqueness of the solution to \eqref{e1.1}, it is
sufficient to prove uniqueness of the solution to 
\eqref{e1.4} in the ball $B(0,R):=\{z_\ve: ||z_\ve||\leq 
R\}$. 
If \eqref{e1.4} has two solutions, say $z$ and $v$,
then their difference $p:=z-v$ solves the equation
$p=-A_\ve^{-1}[K(z)-K(v)]$, where
$$K(z):=\int_0^1(1-s)F^{\prime \prime}(y+sz)z\, z\, ds $$ is
the remainder term in the Taylor formula
$F(u_\ve)-F(y)=Az+K(z)$. If $||z||\leq R$ and $||v||\leq R$,
then one has \be\label{e2.3} ||K(z)-K(v)||\leq
\int_0^1ds(1-s)(M_3 R^2 s+2M_2R)||p||= (\frac{M_3
R^2}6+M_2R)||p||. \ee Thus, using \eqref{e2.5}, one gets:
\be\label{e2.4} ||p||\leq q
||p||,\quad q:=\frac c \ve (\frac{M_3R^2}6+M_2R). 
\ee 
Take $R= \frac {\ve(1-\rho)}{cM_2}$ (see \eqref{e2.2}). 
Then $q\leq 1-\rho+\ve \frac
{M_3(1-\rho)^2}{6cM_2^2}<1$ if $\ve$ is sufficiently small.
Thus, $p=0$ if $\ve$ is sufficiently small.
 \refT{2.1} is proved. \end{proof}

\section{An example}\label{S:3} Consider equation
\eqref{e1.6}. Let $D\subset\R^n$ be a bounded domain, $n=3$,
and
$$F(u):=\int_D g(x,s)u^3(s)ds:=Gu^3,\quad f=0,\quad
F'(u)\psi=3\int_D g(x,s) u^2(s)\psi(s)ds.
$$
 Then $y(s)=0$, $F(y)=0$, $F'(y)=0$, and
$$\|(F'(y)+\ve)^{-1}\|=\frac{1}{\ve},$$ 
so that \eqref{e1.5} holds. We took $f=0$ for simplicity.  

Let $\|u\|_{H^1(D)}:=\|u\|_1$, $\|u\|_{L^2(D)}:=\|u\|_0$.
Let us check that equation \eqref{e1.6} 
with $w=\ve h$, where $\|h\|_1=1$, $h$ is otherwise 
arbitrary, has a unique
solution $u_\ve$ for any $\ve\in(0,\ve_0)$.
Write \eqref{e1.6} as \be\label{e3.1}
  u_\ve=-\frac{1}{\ve} F(u_\ve)+\ve h:=T(u_\ve). \ee
Let us check that $TB_1(R)\subset B_1(R):=\{u:\|u\|_1\leq 
R\}$, and $T$ is a contraction mapping on $B(R)$, where 
$R=\ve^{2/3}$.
As a Hilbert space we take $H=H^1(D)$, the Sobolev space.

If $n=3$ then $\|u\|_{L^6(D)}\leq c\|u\|_{H^1(D)}$ by the
embedding theorem, so $u^3\in L^2(D)$ and 
$$\|Gu^3\|_{H^2(D)}=
\|\int_Dg(x,s)u^3(s)ds\|_{H^2(D)}\leq 
c\|u^3\|_0
$$ 
by the
properties of the Newtonian potential. Thus
$$\|F(u)\|_{H^2(D)}\leq c\|u\|^3_{L^6(D)}\leq
c_1\|u\|^3_{H^1(D)},
$$
 where by $c$ and $c_1$ various
positive constants, independent of $u$, are denoted. One
has $\|h\|_1=1$, and $\|T(u)\|_1\leq \frac{c_1}{\ve}
\|u\|^3_1+\ve$. If $\|u\|_1\leq R$, then $\|T(u)\|_1\leq R$
provided that 
\be\label{e3.2}
  \frac{c_1}{\ve} R^3+\ve\leq R.  \ee 
Choose
$R=\ve^{2/3}$. Then \eqref{e3.2} holds if $c_1\ve+\ve\leq
\ve^{2/3}$. This inequality holds for all $\ve\in(0,\ve_0)$
if $\ve_0>0$ is sufficiently small, namely
$\ve_0<(1+c_1)^{-3}$. 

Thus $T: B_1(R)\to B_1(R)$ if $R=\ve^{2/3}$
and $\ve\in(0,\ve_0)$.

Let us check that $T$ is a contraction on $B_1(R)$. If
$u,z\in B_1(R)$, then one has
$$\|T(u)-T(z)\|_1 =\frac{1}{\ve} \|F(u)-F(z)\|_1
\leq \frac{1}{\ve}\|G(u^3-z^3)\|_1.$$
Furthermore,
$$\|G(u^3-z^3)\|_1 \leq c\|u^3-z^3\|_0
 \leq c\|u-z\|_0 (\|u\|^2_{L^4(D)}+\|z\|^2_{L^4(D)})\leq 
c_2\|u-z\|_1 R^2.
$$
Thus, with $R=\ve^{2/3}$, one has:
$$\|T(u)-T(z)\|_1 \leq \frac{c_2R^2}{\ve} \|u-z\|_1
  \leq c_2\ve^{1/3}\|u-z\|_1.
$$
 If $\ve\in(0,\ve_0)$ and
$\ve^{1/3}_0 c_2<1$, then $T$ is a contraction map on $B(R)$
in the space $H^1(D)$.

Therefore we have proved the following:

\begin{theorem}\label{T:3.1} Assume $\|h\|_1=1$ and
$\ve\in(0,\ve_0)$, where $\ve_0$ is sufficiently small. Then
equation \eqref{e3.1} has a unique solution $u_\ve$ and
$\lim_{\ve\to 0}\|u_\ve\|_1=0$. \end{theorem}

\end{document}